\documentclass{PoS}

\title{Decays of doubly charmed meson molecules}

\ShortTitle{Decays of doubly charmed meson molecules}

\author{\speaker{R. Molina}\\
        Research Center for Nuclear Physics (RCNP), Osaka University, \\
Ibaraki, Osaka 567-0047, Japan\\
        E-mail: \email{molina@rcnp.osaka-u.ac.jp}}
\author{A. Hosaka\\
        Research Center for Nuclear Physics (RCNP), Osaka University, \\
Ibaraki, Osaka 567-0047, Japan\\
        E-mail: \email{hosaka@rcnp.osaka-u.ac.jp}}
\author{H. Nagahiro\\
        Department of Physics, Nara Women's University \\
        E-mail: \email{nagahiro@cc.nara-wu.ac.jp}}

\abstract{The interaction between pseudoscalar and/or vector mesons can be studied using hidden gauge Lagrangians. In this framework, the interaction between charmed mesons has been studied. Furthermore, doubly charmed states are also predicted. These new states are near the $D^*D^*$ and $D^*D^*_s$ thresholds, 
and have spin-parity $J^P=1^+$. 
 We evaluate the decay widths of these states, named as $R_{cc}(3970)$ and $S_{cc}(4100)$ (with strangeness), and obtain $44$ MeV for the non-strangeness, and $24$ MeV for the doubly charm-strange state. Essentially, the decay modes are $DD_{(s)}\pi$ and $DD_{(s)}\gamma$, being the $D\pi$ and $D\gamma$ emitted by one of the $D^*$ meson which forms the molecule.}

\FullConference{XV International Conference on Hadron Spectroscopy-Hadron 2013\\
		4-8 November 2013\\
		Nara, Japan }

\begin{document}
\section{Introduction}
Recently, the LHCb has measured the quantum numbers of the X(3872) as $1^{++}$  \cite{lhcb2013}. This result rules out  the X(3872) to be a charmonium state, favoring the molecular interpretation \cite{olsen}. In addition, several authors have discussed whether some of the other observed XYZ particles can be described in terms of molecules \cite{daniaxial,xyz,xyzother}.  Some of the reasons on why these states cannot be accomodated into $c\bar{c}$ are the unusually high decay rates into $(\rho, \omega$ or$ \phi )J/\psi$ \cite{olsen}. Also, charged states $Z_c$ and decays between them are observed \cite{belle2013}.  

Using hidden gauge Lagrangians combined with unitarity in coupled channels, some of the observed states which are near the open charm thresholds, are well described in terms of two-meson molecules \cite{daniaxial,xyz}. Moreover,  two-meson bound states of $D^{*}D^{*}$ or $D^{*}D^{*}_s$ are dynamically generated \cite{exotic}.
Those doubly charmed mesons form a charged isospin singlet and doublet, they are called $R_{cc}^+(3970)$ and  $S_{cc}^{+(+)}(4100)$, for the non-strangeness and strangeness one respectively. Doubly charm states with the same quantum numbers have also been found in \cite{valcarce} from solving the scattering problem of two $D$-mesons with the interaction provided by the chiral constituent quark model. Theoretically, tetraquark structure has been also discussed \cite{cui,marina,Hyodo:2012}.
 In this talk, in order to explore further the internal structure of these states, we study the decays of these states in detail.  
\section{Decay modes of doubly charm states}
\begin{figure}
\begin{center}
 {\begin{tabular}{@{}cc@{}}
\includegraphics[width=5cm]{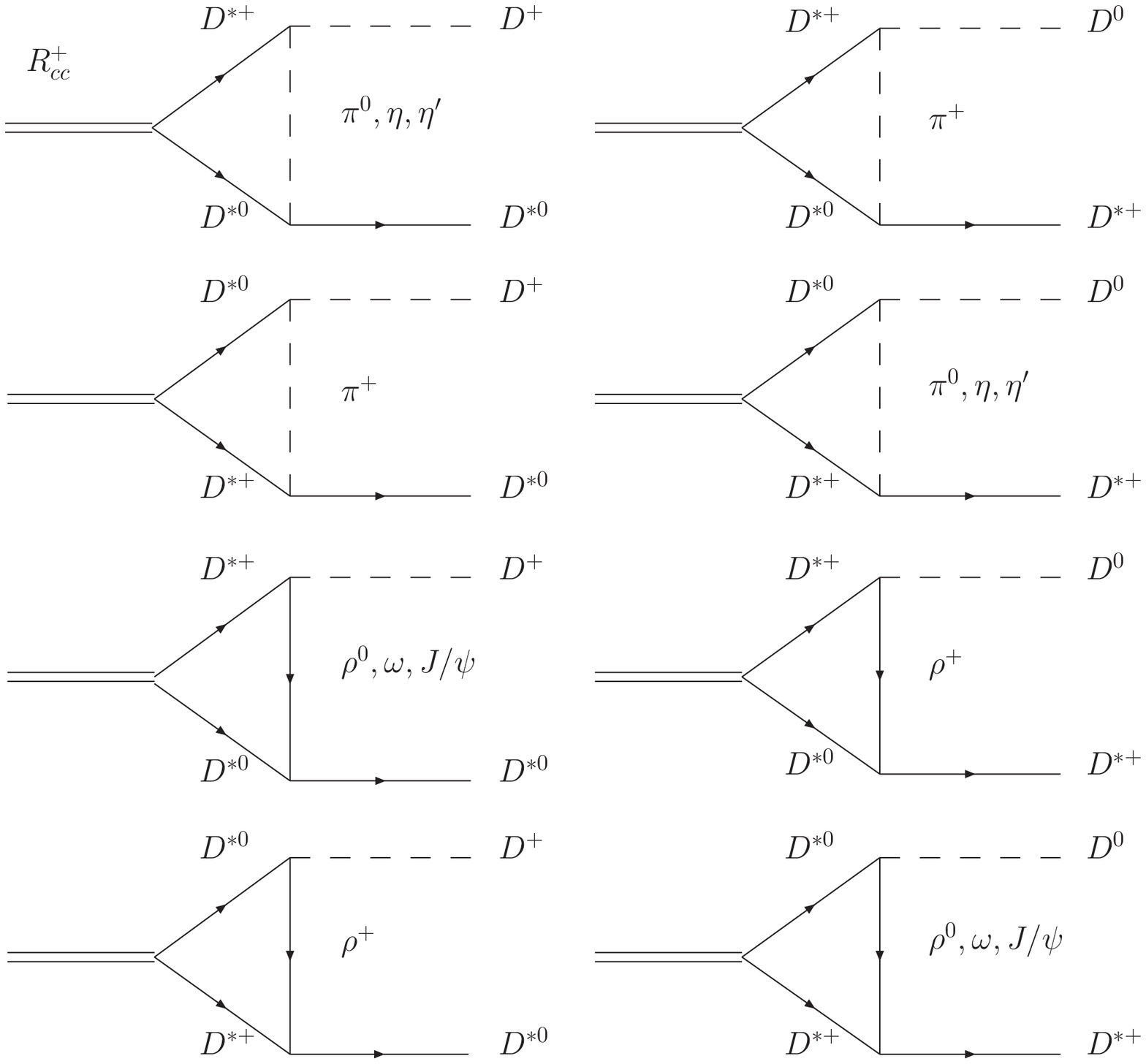}&\includegraphics[width=5cm]{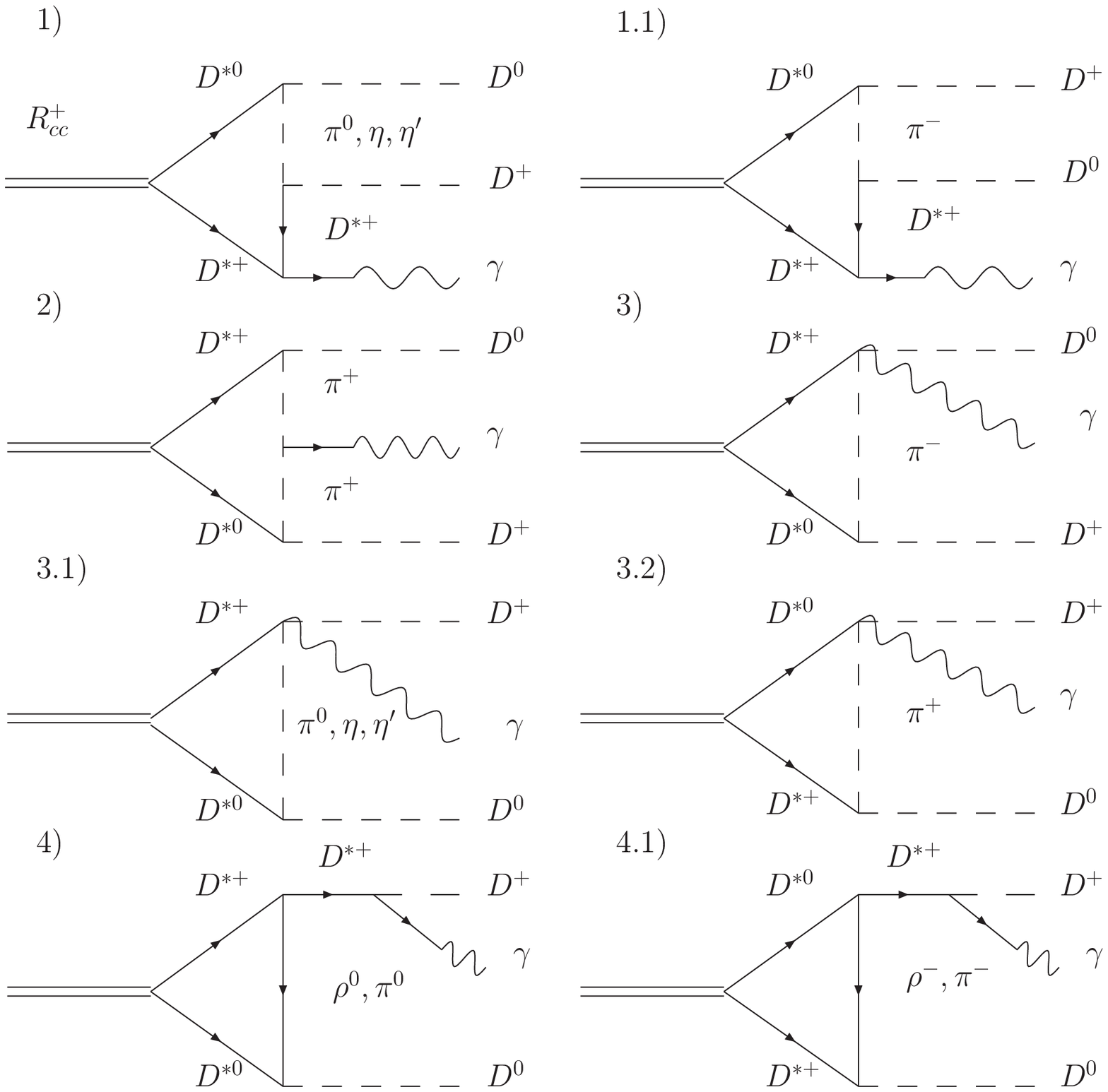}
\label{fig:3} 
\end{tabular}
}\end{center}
\caption{Left: Feynman diagrams evaluated in the decay $R_{cc}\to DD^*$. Right: Diagrams for the $R_{cc}^+\to D^0D^+\gamma$ decay through one loop.}
\end{figure}

The two $D^*$ mesons can form a molecular state of spin and parity $J^P = 1^+$ when they are dominated by an s-wave state. Due to these quantum numbers, it cannot decay into $D\bar{D}$. Strong and radiative decays of the doubly charm states occur through $DD^{*}_{(s)}$ (or $D_{(s)}D^{*}$) which subsequently go to three body states via $D^* \to \pi D$ or $ D \gamma$. Direct decays into three-body states, $DD\gamma$, are also evaluated, but they are small as compared to the above processes going through two bodies. The set of Feynman diagrams considered are depicted in Fig. \ref{fig:3}.
The $R_{cc}\to DD^*_{(s)}$ transition can be reached through anomalous couplings $VVP$ with pseudoscalar or vector meson exchange. The Lagrangians needed to evaluate the decay width to $DD^*_{(s)}$ are \cite{bando},
\begin{eqnarray}
&\mathcal{L}_{PPV}=-ig\langle V^\mu [P,\partial_\mu P]\rangle\ ,\hspace{1cm}\mathcal{L}_{3V}=ig\langle (V^\mu \partial_\nu V_\mu -\partial_\nu V_\mu V^\mu)
V^\nu)\rangle&\label{lag4}\nonumber\\
&\mathcal{L}_{VVP}=\frac{G'}{\sqrt{2}}\epsilon^{\mu\nu\alpha\beta}\langle\partial_\mu V_\nu \partial_\alpha V_\alpha P\rangle&\label{lagan}\ ,
\end{eqnarray}
with $e$ the unit electronic charge, $G' = 3g'^2/(4\pi^2f)$, $g' = -G_V M_\rho/(\sqrt{2}f^2)$, $G_V=f/\sqrt{2}$ and $g=M_V/2f$. The constant $f$ is the pion decay constant $f=93\ MeV$, $Q=diag(2,-1,-1,1)/3$ and $M_V$ is the mass of the vector meson. The $P$ and $V$ matrix contain the 15-plet of the pseudoscalars and vectors respectively in the physical basis. In \cite{exotic} the uncertainties related with the SU(4) breaking of the coupling $g$ are studied, considering both heavy and light couplings.
 These decays come through one loop which involves an integral which is logarithmically divergent, however this divergence is related to the vertex that couples the resonance to the two-meson molecular states, and is also present in the two-meson loop function, $G$, when those states are dynamically generated \cite{exotic}. Thus, the same value of the cutoff needed to obtain these states at their masses \cite{exotic} is used to evaluate the integral involved in the decays in the one-loop diagrams of Fig. \ref{fig:3}. Once set the cutoff, one has a fixed mass and coupling of the bound state to the two-meson component, $g_R$. Since these three magnitudes are related, there is only one free parameter in the calculation, the cutoff $q_{\mathrm{max}}$, and performing variations of this parameter one has an idea of the uncertainties in the decay widths. This is reflected in the errors of the widths, where $15$\% variations around its central value, $750$ MeV, have been considered. 

The diagrams included in the evaluation of the radiative decay of doubly charmed meson molecules,  $R_{cc}\to DD\gamma$ are depicted in Fig. \ref{fig:3} (right panel), where only non-vanishing diagrams are shown.
\section{Results}
 The results are shown in Table \ref{tab:res1}.
 We observe that the total widths of the doubly charmed states are $(44\pm 12)$, $(24\pm 8)$, and $(24\pm 8)$ MeV for the $R^+_{cc}$, $S^+_{cc}$ and $S^{++}_{cc}$ respectively, giving both channels (ex. $D^0D^{*+}$ and $D^+D^{*0}$ for the $R^+_{cc}$) the same contribution to the width. The direct diagrams with three/four propagators of Fig. \ref{fig:3}, type $1)$, $2)$ and $3)$, lead to a very small width of the order of few KeV in the case of the $R_{cc}^+(3970)$ and $S_{cc}^+(4100)$ and $0.13$ KeV for the doubly charge state, $S_{cc}^{++}(4100)$. 
\section{Conclusions}
We have considered the possible decay modes of the doubly charmed molecules, $R_{cc}(3970)$ and $S_{cc}(4100)$, and evaluated partial decay widths to $DD_{(s)}\pi$ and $DD_{(s)}\gamma$. We find that the main source of these decays come from the decay of a $D^*_{(s)}$ meson into $D_{(s)}\pi$ or $D_{(s)}\gamma$. These decays are mediated by the exchange of one meson, vector or pseudoscalar, between the $D^*D^*_{(s)}$ pair of the molecule. The largest width comes from $\rho$, $\pi$ and $\omega$ exchange (decreasing order) for the $R_{cc}(3970)$. Since they are not $q\bar{q}$, having a pair of $cc$ and doubly charged, these mesons are under challenge for experiments. Hopefully, they could be observed by the LHCb or Belle.
{\tiny{\begin{table}
 {\begin{tabular}{@{}cccccc@{}}\hline\hline
{\it State}& {\it Channel k}&$\Gamma^k$ [MeV] &{\it Channel j}&$\Gamma^k_j$ [MeV]&$\Gamma_{\mathrm{tot}}$ [MeV]\\\hline
$R^+_{cc}(3970)$&\multicolumn{5}{c}{{\it{Hadronic decays}}}\\
&$D^0D^{*+}$&$22\pm 6$&$D^0(D^{+}\pi^0)$&$7\pm 2$ & $44\pm 12$\\
&&&$D^0(D^{0}\pi^+)$& $15\pm 4$&\\
 &$D^+D^{*0}$&$22\pm 6$ &$D^+(D^0\pi^0)$&$14\pm 4$&\\
&\multicolumn{5}{c}{{\it{Radiative decays}}}\\
&$D^+D^{*0}$& &$D^+(D^0\gamma)$& $8\pm 2$&\\
&$D^0D^{*+}$&&$D^0(D^+\gamma)$&$0.4\pm 0.2$& \\
&& &$D^0D^+\gamma$&$(2\pm 1)\times10^{-3}$ &\\
 &&&$D^{*0}D^+\gamma$& $ (0.03\pm 0.01)\times10^{-3}$& \\
&& &$D^{*+}D^0\gamma$&$(0.5\pm 0.2)\times 10^{-3}$ & \\
 \hline
$S^+_{cc}(4100)$&\multicolumn{5}{c}{{\it{Hadronic decays}}}\\
&$D^+_sD^{*0}$&$12\pm 4$&$D^+_s(D^{0}\pi^0)$&$7\pm 2$&$24\pm 8$ \\
&$D^0D^{*+}_s$&$12\pm 4$&-&-&\\
&\multicolumn{5}{c}{{\it{Radiative decays}}}\\
&$D^0D^{*+}_s$&&$D^0(D^+_s\gamma)$&$11\pm 4$&\\
&$D^+_sD^{*0}$&&$D^+_s(D^0\gamma)$& $5\pm 2$&\\
&&&$D^0D^+_s\gamma$&$(2\pm 1)\times10^{-3}$ &  \\
&&&$D^{*0}D^+_s\gamma$&$(0.3\pm 0.1) \times 10^{-3}$ & \\
 &&&$D^{*+}_sD^0\gamma$&$(4\pm 1)\times 10^{-3}$ &\\
\hline
$S^{++}_{cc}(4100)$&\multicolumn{5}{c}{{\it{Hadronic decays}}}\\
&$D^+_sD^{*+}$&$12\pm 4$&$D^+_s(D^{+}\pi^0)$&$4\pm 1$& $24\pm 8$\\
&&&$D^+_s(D^{0}\pi^+)$&$8\pm 3$& \\
&$D^+D^{*+}_s$&$12\pm 4$&-&-&\\
&\multicolumn{5}{c}{{\it{Radiative decays}}}\\
&$D^+D^{*+}_s$&&$D^+(D^+_s\gamma)$&$11\pm 4$&\\
&$D^+_sD^{*+}$&&$D^+_s(D^+\gamma)$& $0.2\pm 0.1$&\\
&&&$D^+D^+_s\gamma$&$(1.3\pm 0.1)\times10^{-4}$ & \\
&&&$D^{*+}D^+_s\gamma$&$(0.3\pm 0.1)\times 10^{-3}$ & \\
 &&&$D^{*+}_sD^+\gamma$&$(0.3\pm 0.1)\times 10^{-3}$ &\\
 \hline\hline
\end{tabular}
\caption{Total and partial decay widths of the different decay modes of the doubly charmed states. }}
\label{tab:res1}
\end{table} }}


\end{document}